\documentclass{article}
\usepackage[T1]{fontenc} 
\usepackage[utf8]{inputenc} 
\usepackage{ismir,amsmath,cite,url}
\usepackage{graphicx}
\usepackage{color}
\usepackage{multirow}
\usepackage{array}
\usepackage{subcaption}
\usepackage{enumitem}

\usepackage{enumitem}
\setitemize{noitemsep,topsep=0pt,parsep=0pt,partopsep=0pt}

\usepackage{amsfonts}
\usepackage{xcolor}

\title{JukeDrummer: Conditional Beat-aware Audio-Domain Drum Accompaniment Generation via Transformer VQ-VAE}

\threeauthors
  {Yueh-Kao Wu} {Academia Sinica \\ {\tt yk.lego09@gmail.com}}
  {Ching-Yu Chiu} {National Cheng Kung University \\ {\tt x2009971@gmail.com}}
  {Yi-Hsuan Yang} {Taiwan AI Labs \\ {\tt yhyang@ailabs.tw}}

\def\authorname{Y.-K. Wu, C.-Y. Chiu, and Y.-H. Yang}

\usepackage[bookmarks=false,hidelinks]{hyperref}
\begin{document}

\maketitle
\begin{abstract}

This paper proposes a model that generates a drum track in the audio domain to play along to a user-provided drum-free recording. Specifically, using paired data of drumless tracks and the corresponding human-made drum tracks, we train a Transformer model to  improvise the drum part of an unseen drumless recording. We combine two approaches to encode the input audio. First, we train a vector-quantized variational autoencoder (VQ-VAE) to represent the input audio with discrete codes, which can then be readily used in a Transformer. Second, using an audio-domain beat tracking model, we compute beat-related features of the input audio and use them as embeddings in the Transformer. Instead of generating the drum track directly as waveforms, we use a separate VQ-VAE to encode the mel-spectrogram of a drum track into another set of discrete codes, and train the Transformer to predict the sequence of drum-related discrete codes. The output codes are then converted to a mel-spectrogram with a decoder, and then to the waveform with a vocoder. We report both objective and subjective evaluations of variants of the proposed model, demonstrating that the model with beat information generates drum accompaniment that is rhythmically and stylistically consistent with the input audio.

\end{abstract}

\section{Introduction}\label{sec:introduction}



Deep generative models for musical audio generation have witnessed great progress in recent years \cite{engel2017neural,engel2019gansynth,donahue19iclr,lattner2019highlevel,ungan2020,ddsp,dhariwal2020jukebox,karasinger}. 
While models for generating symbolic music such as MIDI \cite{payne2019musenet,donahue2019lakhnes,chen2020automatic,wang20pianotree} 
or musical scores \cite{score-transformer} focus primarily on the \emph{composition} of musical content, an audio-domain music generation model deals with \emph{sounds} and thereby has extra complexities related to timbre and audio quality. 
For example, while a model for generating symbolic guitar tabs can simply consider a guitar tab as a sequence of notes 
\cite{chen2020automatic}, a model that generates audio recordings of guitar needs to determine not only the underlying sequence of notes but also the way to render (synthesize) the notes into sounds. 
Due to the complexities involved, research on deep generative models for musical audio begins with the simpler task of synthesizing individual musical notes \cite{engel2017neural,engel2019gansynth,donahue19iclr}, dispensing the need to consider the composition of notes.
Follow-up research \cite{lattner2019highlevel,ungan2020,ddsp} extends the capability to generating musical passages of a single instrument.
The Jukebox model \cite{dhariwal2020jukebox} proposed by OpenAI greatly advances the state-of-the-art by being able to, quoting their sentence, ``generate high-fidelity and diverse songs with coherence up to multiple minutes.''
Being trained on a massive collection of audio recordings with the corresponding lyrics but not the symbolic transcriptions of  music, 
Jukebox generates multi-instrument music as raw waveforms directly without an explicit model of the underlying sequence of notes.



This work 
aims to improve upon Jukebox in 
two aspects. 
First, 
the backbone of  Jukebox is a hundred-layer Transformer \cite{vaswani2017attention,sparsetransformer} with billions of parameters that are trained with 1.2 million songs on hundreds of NVIDIA V100 GPUs for weeks at OpenAI, which is hard to reproduce elsewhere.
Inspired by a recent Jukebox-like model for singing voice generation called KaraSinger \cite{karasinger}, we instead build a light-weight model with only 25 million parameters by working on Mel-spectrograms instead of raw waveforms.
Our model is trained with only 457 recordings on a single GeForce GTX 1080 Ti GPU for 2 days.

Second, and more importantly, instead of a fully autonomous model that makes a song from scratch with various instruments, we aim to build a model that can work cooperatively with human, allowing the human partner to come up with the musical audio of \emph{some} instruments as input to the model, and generating in return the musical audio of \emph{some other} instruments to accompany and to complement the user input, completing the song together.
Such a model can potentially contribute to human-AI co-creation in songwriting 
\cite{huang20ismir} and enable new applications.

In technical terms, our work enhances the controllability of the  model by allowing its generation to be steered on a user-provided audio track. 
It can be viewed as an interesting sequence-to-sequence problem where the model creates a ``target sequence'' of music that is to be played along to the input ``source sequence.'' Besides requirement on audio quality, the coordination between the source and target sequences in terms of musical aspects such as style, rhythm, and harmony is also of central importance.

We note that, for controllability and the intelligibility of the generated singing, both Jukebox \cite{dhariwal2020jukebox} and KaraSinger \cite{karasinger} have a lyrics encoder that allows 
their generation to be steered on textual lyrics. While being technically similar,
our \emph{accompaniment generation} task (``audio-to-audio'') is different from the lyric-conditioned generation task (``text-to-audio'') in that the latter does not need to deal with the coordination between two audio recordings.


Specifically, we consider a \emph{drum accompaniment generation} problem in our implementation, using a ``drumless'' recording as the input and generating as the output a drum track that involves the use of an entire drum kit.
We use this as an example task to investigate the audio-domain accompaniment generation problem out of the following reasons.
First, datasets used in musical source separation \cite{musdb18} usually consist of an isolated drum stem along with stems corresponding to other instruments. 
We can therefore easily merge the other stems to create paired data of drumless tracks and drum tracks as training data of our model. (In musical terms, drumless, or ``Minus Drums'' songs are recordings where the drum part has been taken out, which corresponds nicely to our scenario.)
Second, we suppose a drum accompaniment generation model can easily find applications in songwriting \cite{deruty22tismir}, as it allows a user (who may not be familiar with drum playing or beat making) to focus on the other non-drum tracks.
Third, audio-domain drum accompaniment generation poses interesting challenges as the model needs to determine not only the \emph{drum patterns} but also the \emph{drum sounds} that are supposed to be, respectively, rhythmically and stylistically consistent with the input.
Moreover, the generated drum track is  expected to follow a steady tempo, which is a basic requirement for a human drummer.
We call our model the ``JukeDrummer.''

\begin{figure}[t]
    \centering
    \includegraphics[width=.56\linewidth]{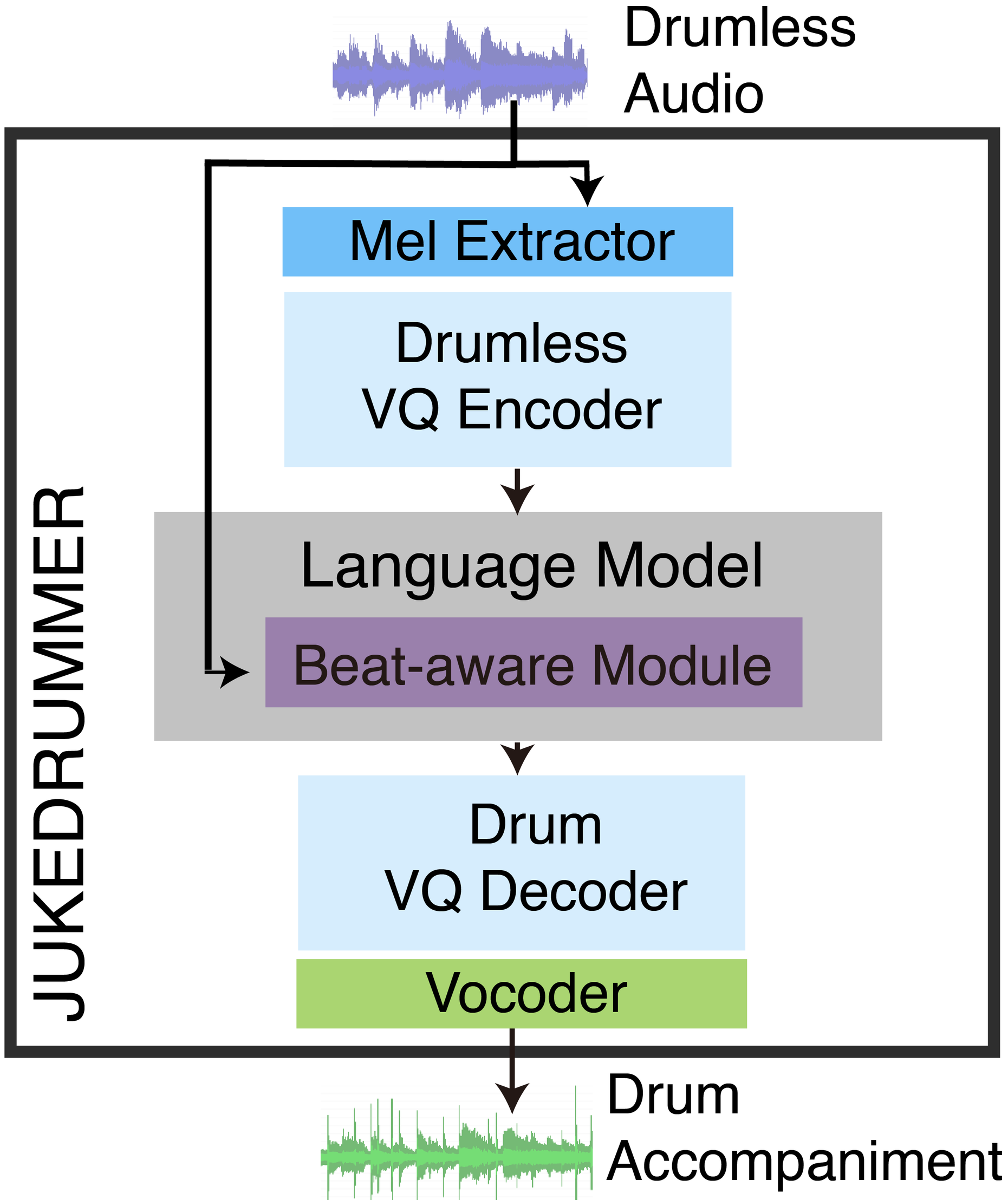}
    \caption{Diagram of the proposed JukeDrummer model for the inference stage. The training stage involves learning additional Drum VQ Encoder and Drumless VQ Decoder (see Figure \ref{fig:2VQ}) that are not used at inference time.}
    \label{fig:diagram}
\end{figure}

As depicted in Figure \ref{fig:diagram}, the proposed model architecture contains an ``audio encoder'' (instead of the original text encoder \cite{dhariwal2020jukebox,karasinger}) named the \emph{drumless VQ encoder} that takes a drum-free audio as input.
Besides, we experiment with different ways to capitalize an audio-domain beat and downbeat tracking model
proposed recently \cite{spl21chiu} in a novel \emph{beat-aware module} that extracts \emph{beat-related information} from the input audio, so that the language model for generation (i.e., the Transformer) is better informed of the rhythmic properties of the input.
The specific model \cite{spl21chiu} was trained on drumless recordings as well, befitting our task.
We extract features from different levels, including low-level tracker embeddings, mid-level activation peaks, and high-level beat/downbeat positions, and investigate which one benefits the generation model the most.


Our contribution is four-fold.
First, to our best knowledge, this work represents 
the first attempt to drum accompaniment generation of a full drum kit given drum-free mixed audio.
Second, we develop a light-weight audio-to-audio Jukebox variant that takes an input audio of up to 24 seconds as conditioning and generates accompanying music in the domain of Mel-spectrograms (Section \ref{sec:methods}).
Third, we experiment with different beat-related conditions in the context of audio generation (Section \ref{sec:BeatCondEmbed}).
Finally, we report objective and subjective evaluations demonstrating the effectiveness of the proposed model (Sections \ref{sec:exp} \& \ref{sec:exp2}).\footnote{We share our code and checkpoint at: \url{https://github.com/legoodmanner/jukedrummer}. Moreover, we provide audio examples at the following demo page: \url{https://legoodmanner.github.io/jukedrummer-demo/}}.

\section{Background}\label{sec:related_works}

\subsection{Related Work on Drum Generation}
\label{sec:related_works:drum}

Conditional drum accompaniment generation has been studied in the literature, but only in the symbolic domain \cite{dahale21ismir-lbd,makris22evomusart}, to the best of our knowledge. 
Dahale \emph{et al.} \cite{dahale21ismir-lbd} used a Transformer encoder to generate an accompanying symbolic drum pattern of 12 
bars given a four-track, melodic MIDI passage.
Makris \emph{et al.} \cite{makris22evomusart} adopted instead a sequence-to-sequence architecture with a bi-directional long short-term memory (BLSTM) encoder extracting information from the melodic input and a Transformer decoder generating the drum track for up to 16 bars in MIDI format.
While symbolic-domain music generation has its own challenges, it differs greatly from the audio-domain counterpart studied in this paper, for it is not about generating sounds that can be readily listened to by human.




Related tasks that have been attempted in the literature with deep learning include symbolic-domain generation of a monophonic drum track (i.e., kick drum only) of multiple bars \cite{lattner2019highlevel},  symbolic-domain drum pattern generation \cite{DBLP:conf/icml/GillickREEB19,drumVAE,alain2020deepdrummer,tokui2020gan}, 
symbolic-domain drum track generation as part of a multi-track MIDI \cite{musegan,simon2018learning,8614139,ren2020popmag}, audio-domain one-shot drum hit generation \cite{drumgan,9053128,aouameur2019neural,drysdale2020adversarial,drysdale21ismir-lbd}, audio-domain generation of drum sounds of an entire drum kit of a single bar \cite{tomczak20mm}, and audio-domain drum loop generation \cite{hung21ismir}.
Jukebox \cite{dhariwal2020jukebox} generates a mixture of sounds that include drums, but not an isolated drum track.
By design, Jukebox does not take any input audio as a condition and generate accompaniments.

\begin{figure}
    \centering
    \includegraphics[width=.81\linewidth]{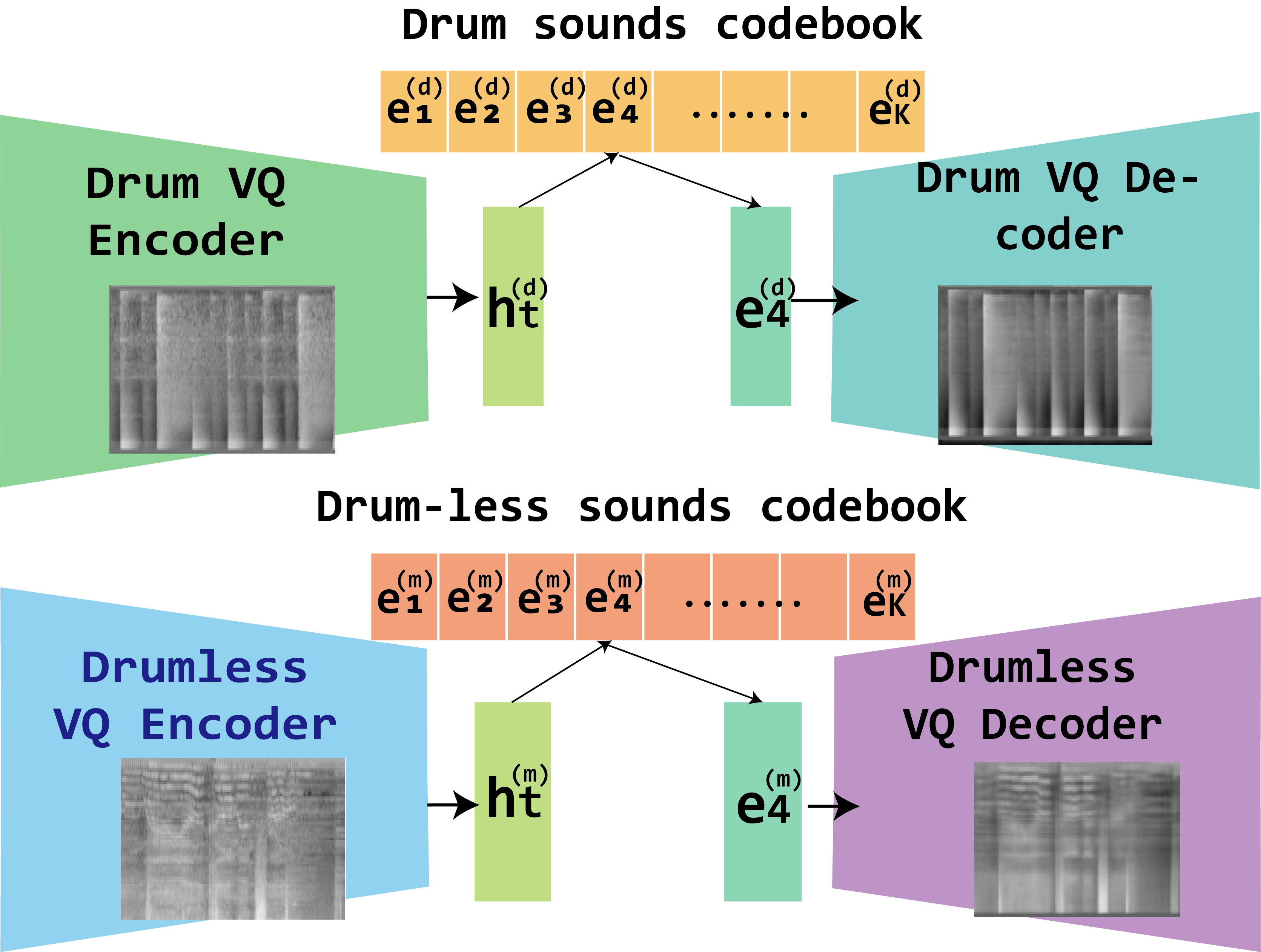}
    \caption{We use separate VQ-VAEs for the drumless and drum tracks, both operating on the Mel-spectrograms.}
    \label{fig:2VQ}
\end{figure}

\subsection{The Original Jukebox model}

The main architecture of Jukebox  \cite{dhariwal2020jukebox} is composed of two components: a multi-scale vector-quantized  variational autoencoder (VQ-VAE) \cite{oord2017neural,kaiser18a,razavi2019generating,polyak21interspeech,walker21arxiv}
and an autoregressive Transformer decoder \cite{vaswani2017attention,sparsetransformer}.
The \textbf{VQ-VAE} is for converting a continuous-valued raw audio waveform 
into a sequence of so-called discrete \emph{VQ codes}, while the \textbf{Transformer} establishes a language model (LM) of the VQ codes capable of generating new code sequences.

Specifically, VQ-VAE consists of an encoder and a decoder, referred to as the VQ-VAE encoder $\mathcal{E}$ and VQ-VAE decoder $\mathcal{D}$  below, respectively.
Given an audio waveform $\mathbf{x} \in \mathbb{R}^{1 \times N}$, where $N$ denotes the number of audio samples (e.g., $N=\text{1,058,400}$ for a 24-second audio clip sampled at 44.1\,kHz), the VQ-VAE encoder would convert $\mathbf{x}$ into a sequence of latent vectors $\{\mathbf{h}_t\in \mathbb{R}^D,t=1,...,T\}$, where the sequence length $T$ is proportional to $N$, and the VQ-VAE decoder would have to reconstruct $\mathbf{x}$ from the ``vector-quantized'' version of the latent vectors, denoting as $\{\mathbf{h}'_t\in \mathbb{R}^D,t=1,...,T\}$, that is obtained by finding the nearest prototype vector $\mathbf{h}'_t = \mathbf{e}_z$ of each $\mathbf{h}_t$ in a codebook of prototype vectors $\{\mathbf{e}_k\in \mathbb{R}^D ,k=1,...,
K\}$, where $K$ is the size of the codebook. 
Each prototype vector can be regarded as a cluster centroid in the latent space as a result of $K$-means clustering.
The VQVAE encoder/decoder and the codebook are jointed learned by minimizing the reconstruction loss $\| \mathbf{x}_t - 
\mathcal{D}(\mathbf{h}'_t)\|^2_2$, and the commitment loss of the clustering $\| \mathcal{E}(\mathbf{x}_t) - sg(\mathbf{e}_z)\|^2_2$, where $\mathbf{x}_t$ denotes a slice of $\mathbf{x}$, $sg(.)$ the stop-gradient operation, and $\mathbf{h}_t=\mathcal{E}(\mathbf{x}_t)$.

Once the VQ-VAE is trained, the $\mathbf{x}$ can be viewed as a sequence of ``IDs'' $\{z_t\in \mathbb{Z}_{1:K},t=1,...,T\}$, each corresponding to the index of the element of the codebook that is used to represent each slice of $\mathbf{x}$.
The Transformer can then be trained on such sequences of IDs to learn the underlying language, or ``composition rules,''  of the audio codes. 
Once trained, the transformer can be used to generate a novel sequence of codes, which can then be converted into an audio waveform by the VQVAE decoder.

As the waveforms are extremely long, Jukebox actually uses a ``multi-scale'' VQ-VAE that converts a waveform into three levels of codes, and accordingly three Transformers for building the LM at each level \cite{dhariwal2020jukebox}. 
KaraSinger works on Mel-spectrograms but also uses multi-scale VQ-VAE \cite{karasinger}.
We simplify their architecture by working on Mel-spectrograms with only one level of codes instead of multiple levels, as introduced below.

\section{Proposed Methods}\label{sec:methods}
In our task, we are given pairs of audio waveforms, namely a drumless stem $\mathbf{x}^\texttt{m}$ and a drum stem $\mathbf{x}^\texttt{d}$. Our goal is to train a model that can generate $\mathbf{x}^\texttt{d}_*$ given an unseen  $\mathbf{x}^\texttt{m}_*$ from the test set. To achieve so, we propose several extensions of the Jukebox model. While we focus on drum accompaniment generation only, the same methodology may apply equally well to other conditional generation tasks.

\textbf{Two VQ-VAEs.} 
As illustrated in Figure \ref{fig:2VQ}, we build separate VQ-VAEs for $\mathbf{x}^\texttt{m}$ and $\mathbf{x}^\texttt{d}$ using drumless stems and drum stems respectively.
Once trained, the drumless VQ-VAE encoder and the drum VQ-VAE encoder would convert  $\mathbf{x}^\texttt{m}$ and  $\mathbf{x}^\texttt{d}$ into $\{z_t^\texttt{m}\in \mathbb{Z}_{1:K^\texttt{m}},t=1,...,T\}$ and $\{z_t^\texttt{d}\in \mathbb{Z}_{1:K^\texttt{d}},t=1,...,T\}$ separately.
Assuming that the drumless tracks are more diverse, we suggest use a larger codebook size for the drumless sounds codebook than the drum sounds codebook, namely $K^\texttt{m} \ge K^\texttt{d}$.


\begin{figure}
    \centering
    \includegraphics[width=.84\linewidth]{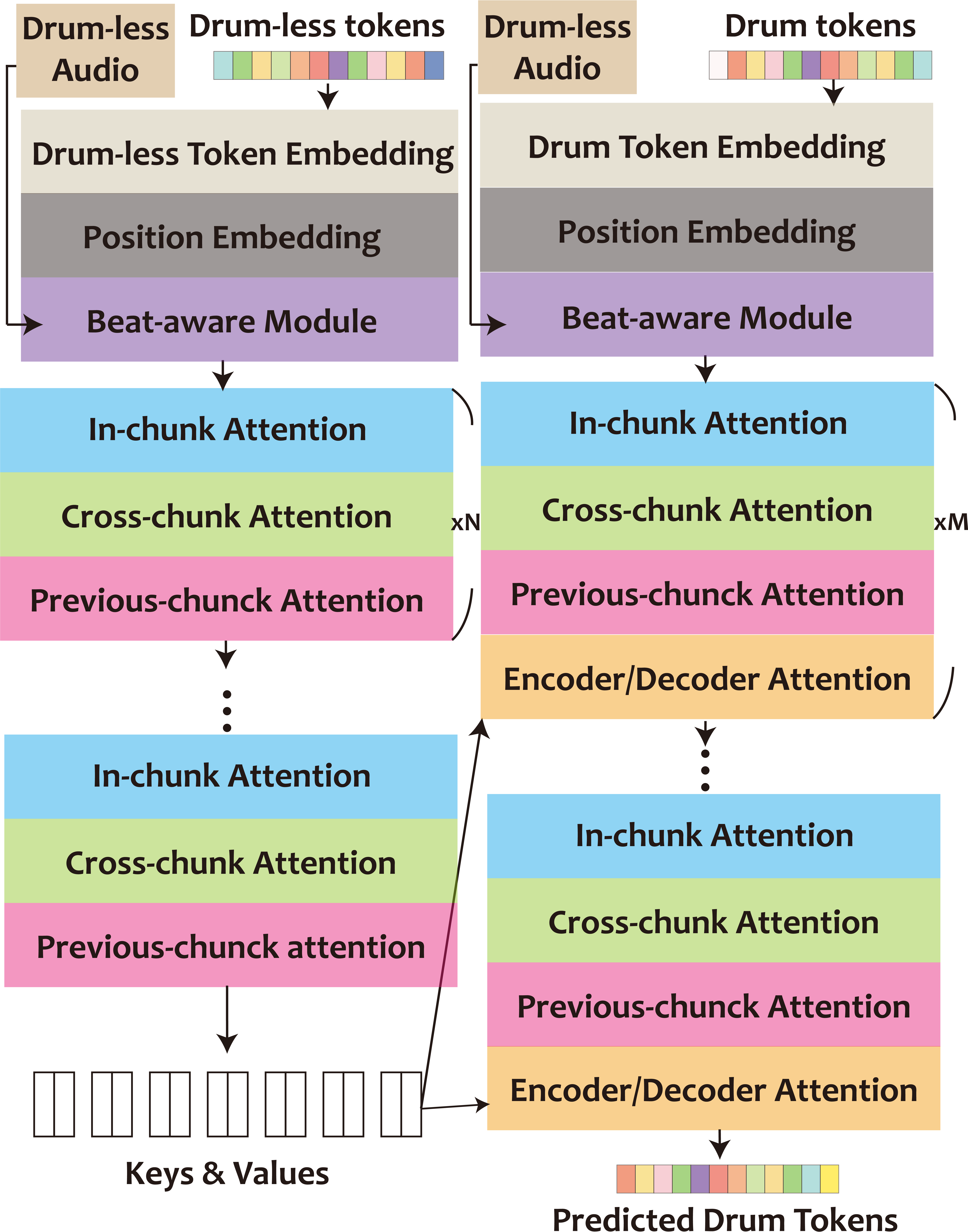}
    \caption{Details of our Transformer encoder/decoder.}
    \label{fig:transformer}
\end{figure}

\begin{figure*}
    \centering
    \includegraphics[width=0.72\textwidth]{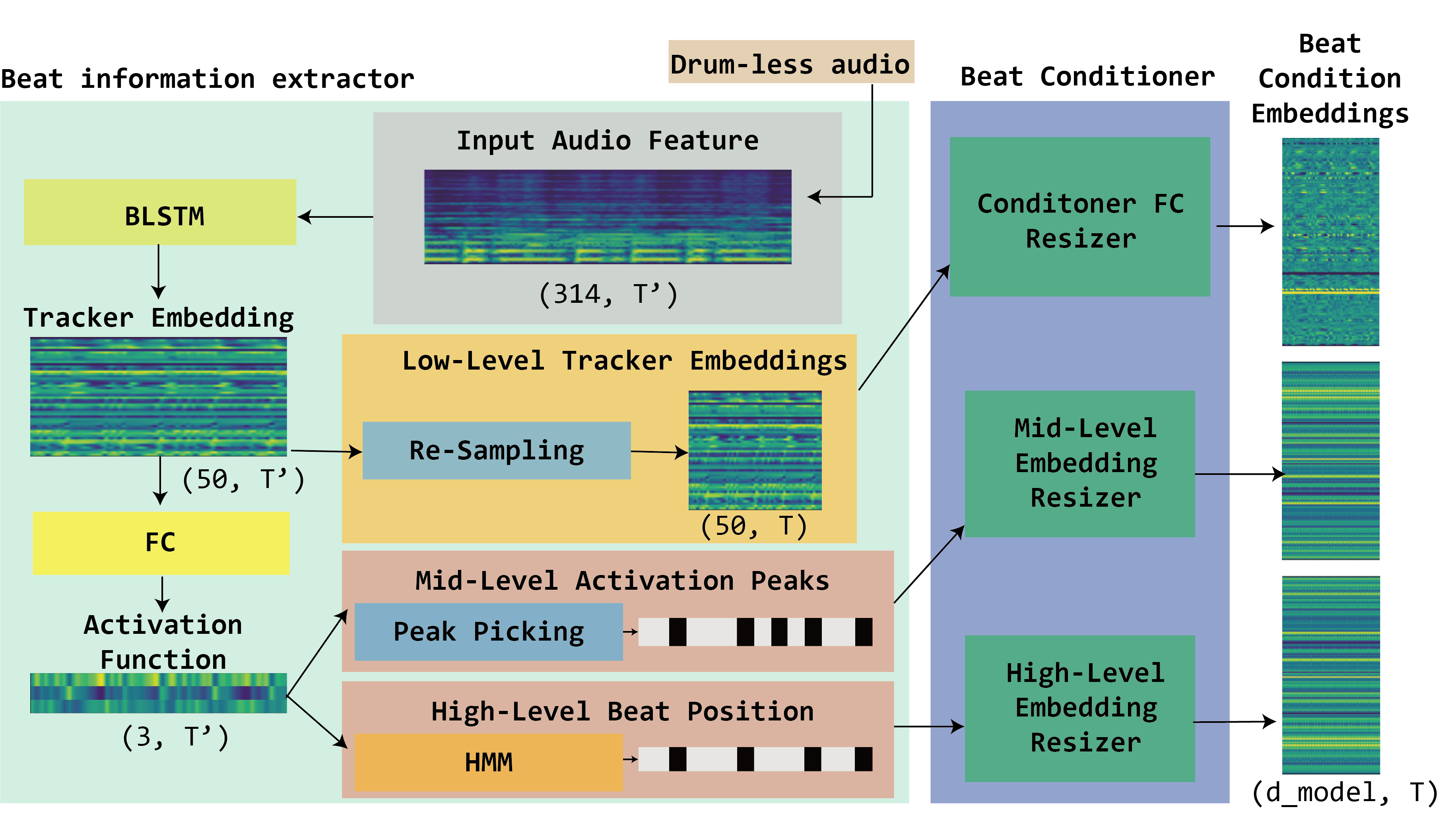}
    \caption{Details of the proposed beat-aware module that extracts beat condition embeddings from the drumless audio.}
    \label{fig:beatinfo}
\end{figure*}

\textbf{Mel VQ-VAE \& Vocoder.} 
To reduce computational cost, we build VQ-VAEs that take Mel-spectrograms as the input and target output.  
Specifically, we use the convolutional blocks of UNAGAN \cite{ungan2020}
to build a pair of VQ-VAE encoder and VQ-VAE decoder that has symmetric architectures (one downsampling and the other upsampling). We omit the details of UNAGAN due to space limit.
Our Transformer accordingly learns the LM for codes of the  Mel-spectrograms.
Once a novel sequence of (drum) codes is generated by the Transformer, we use our (drum) VQ-VAE decoder to convert the codes into a Mel-spectrogram, and then use a neural \emph{vocoder} \cite{hifigan} to convert the Mel-spectrogram into the corresponding waveform.
While the latent vector $\mathbf{h}_t$ (and accordingly $z_t$) corresponds to a slice of waveform in the original Jukebox, here the latent vectors $\mathbf{h}_t^\texttt{m}$ and $\mathbf{h}_t^\texttt{d}$ (and $z_t^\texttt{m}$, $z_t^\texttt{d}$) both correspond to a fixed number of $L$ frames of the short-time Fourier Transform (STFT) while computing the Mel-spectrograms.

\textbf{Seq2seq Transformer.} 
While the Jukebox model uses a Transformer decoder to model the sequences $\{z_t ,t=1,...,T\}$ corresponding to mixtures of sounds, in our case we need a dedicated Transformer encoder to take the sequence $\{z_t^\texttt{m} ,t=1,...,T\}$ corresponding to a drumless audio as the input, and a Transformer decoder to generate the sequence $\{z_t^\texttt{d} ,t=1,...,T\}$ corresponding to the accompanying drum audio as the output, leading to a sequence-to-sequence (seq2seq) architecture.
Following Jukebox, we use the attention mechanism of the Scalable Transformer \cite{sparsetransformer} in our Transformer encoder/decoder. As depicted in Figure \ref{fig:transformer}, to determine its output $z_i^\texttt{d}$ at each $i$, our Transformer decoder uses factorized ``in-chunk,'' ``cross-chunk'' and ``previous-chunk'' attention layers to attend to the drum codes it generates previously, namely $z^\texttt{d}_{<i}=\{z_t^\texttt{d} ,t=1,...,i-1\}$, to maintain the coherence of its output.
A ``chunk'' here is a slice of the corresponding sequence.
Moreover, the Transformer decoder uses ``encoder/decoder attention'' to attend to the final layer of the Transformer encoder to get contextual information from the drumless code sequence $\{z_t^\texttt{m} ,t=1,...,T\}$.
To better synchronize the input and output, position embeddings are used. We refer readers to \cite{dhariwal2020jukebox,sparsetransformer} for details. 



\textbf{Beat-Aware Module.}
Figure \ref{fig:transformer} also shows that we use a novel ``beat-aware module'' (after the position embedding) in both Transformer encoder and decoder. 
We note that the  code $z_t^\texttt{m}$ is trained to provide essential information needed to reconstruct the drumless audio $\mathbf{x}^\texttt{m}_t$, so the information carried by $z_t$ might be \emph{mixed}, covering different musical aspects including timbre, style, rhythm, etc. To supply the Transformer with \emph{clear} rhythm-related information of $\mathbf{x}^\texttt{m}$, we might need such a dedicated beat-aware module.  


As depicted in Figure \ref{fig:beatinfo}, the beat-aware module consists two sub-modules, the ``beat information extractor'' and the ``beat conditioner.''
The former extracts beat-related information per frame 
from $\mathbf{x}^\texttt{m}$ using an existing beat/downbeat tracker \cite{spl21chiu}, while the latter incorporates that beat-related information for each $t$ as a beat condition embedding $\mathbf{c}^\texttt{m}_t \in \mathbb{R}^{d_\text{model}}$ that has the same length $d_\text{model}$ as the token embedding of the Transformers, and adds together the beat condition embedding and token embedding
per $t$ to serve as the input to the subsequent attention layers.


Despite that deep learning-based beat/downbeat trackers such as those proposed by B{\"{o}}ck \emph{et al.} 
\cite{madmom, Bock2016d,Bock2020} have achieved excellent performance for mainstream pop, rock or dance music, their performance and behaviors are influenced by the sound source composition (i.e., drum/non-drum sounds) of their training data \cite{spl21chiu, chiu21eusipco}.
Considering that our input music is drumless, which is quite different from the music that existing common beat/downbeat trackers \cite{Bock2016d, Bock2020} are trained with and trained for, we use in our beat information extractor the ``non-drum tracker'' developed by Chiu \emph{et al.}  \cite{spl21chiu} exclusively for drumless stems.
In Section \ref{sec:BeatCondEmbed}, we describe three ways to extract different levels of beat information feature from the non-drum tracker.

\section{Beat Condition Embedding}\label{sec:BeatCondEmbed}

As shown on the left of Figure \ref{fig:beatinfo}, following the common design \cite{Bock2016d}, the beat/downbeat tracker \cite{spl21chiu} uses BLSTM layers to get 50-dimensional ``tracker embeddings'' from $\mathbf{x}^\texttt{m}$, and then uses fully-connected (FC) layers to compute 3-dimensional ``activation functions'' indicating the likelihood of observing a beat, downbeat, or non-beat at each frame.
Finally, either a simple peak-picking or a 
hidden Markov model (HMM)-based algorithm
\cite{Bock2016d}  can be used to
finalize the beat and downbeat time positions from the activation functions. 
We accordingly investigate extracting features from different stages of this processing pipeline.

\textbf{Low-level (tracker) embeddings.}
We simply use the high-dimensional tracker embeddings, resampling it temporally, pooling them every $L$ frames,
and converting each of them to the desired length $d_\text{model}$ with an FC.



\textbf{High-level beat/downbeat positions.}
From the beat and downbeat time positions estimated by HMM, a frame can be labeled as either a beat, downbeat, or non-beat. We learn three $d_\text{model}$-dimensional embedding vectors corresponding to each case, and represent every $L$ frames with one of the three vectors according to the frame labels.

\textbf{Mid-level activation peaks.}
As the beat/downbeat positions are sparse over time, we consider a ``denser'' version by simply picking the peak positions of the activation functions by  \texttt{scipy.signal.find\_peaks} \cite{scipy} and similarly learning embedding vectors that are assigned to the frames according to whether a frame corresponds to a peak or not.
Such a peak might represent a musical onset. 

\section{Experiment Setup}
\label{sec:expsetup}


Multi-track datasets for research on musical source separation \cite{musdb18} 
usually host recordings that each consists of an isolated drum stem along with stems corresponding to other instruments. 
We can simply take the drum stem as the target drum track, and the summation of the remaining stems as the input drumless track.
In our implementation, we used the multi-track recordings from  three datasets.  
MUSDB18 \cite{musdb18} contains 150  recordings, each of which has four stems corresponding to vocal, drums, bass, and ``others.'' It is commonly used in source separation.
MedleyDB \cite{bittner14ismir} contains 196 multi-track recordings, each of which includes a drum stem along with many other stems. 
MixingSecret \cite{mixingsecret}
has 257 multi-track recordings with various instruments, all including drums. 
We removed the 46 duplicate recordings between MUSDB18 and MedleyDB and the 100 duplicate recordings between MUSDB18 and MixingSerect, leading to 457 recordings to be used in our work. 
We randomly split the recordings into 80\%, 10\%, 10\%  as the  training set, validation set (for parameter tuning), and testing set (for objective and subjective evaluation) at the ``recording-level,'' ensuring that a recording does not appear in different splits. 

All the recordings are sampled at 44.1\,kHz and the stems are all monaural.
Following Jukebox \cite{dhariwal2020jukebox}, we used 24-second audio clips in our work.
We sliced each recording (and accordingly the stems) to 23.8-second audio clips with 50\% temporal overlaps  (as $2^{20}/\text{44100}=\text{23.8}$), and discarded the clips that do not contain any drum sounds. We then computed the Mel-spectrograms of each clip with PyTorch v1.7.1 with a Hann window of 1,024 samples for STFT, a hop size of 256 samples and 80 Mel-filter banks.

To evaluate the performance of the proposed JukeDrummer and validate the effectiveness of model components, we adopted the following  variants in our experiments.\footnote{A reviewer suggested that we should have compared our model with other accompaniment generation models that operate in the symbolic domain such as \cite{dahale21ismir-lbd} and \cite{makris22evomusart}, saying that we can use existing audio samples or drum synthesis models to render their output to audio. However, the problem is that such a symbolic-domain accompaniment generation model also requires its input to be a MIDI file rather than audio.}
\begin{itemize}
    \item \texttt{seq2seq+beat\,(low)}: given a drumless clip $\mathbf{x}^\texttt{m}$, we computed the drumless codes $\{z_t^\texttt{m}\}$ and the low-level beat/downbeat tracker embeddings $\{\mathbf{c}^\texttt{m}_t\}$ as the model input, to predict the drum codes $\{z_t^\texttt{d}\}$ that eventually lead to the generated drum clip $\mathbf{x}^\texttt{d}$, using the proposed seq2seq Transformer.
    \item \texttt{seq2seq+beat\,(mid)}: using the beat condition embeddings computed from the mid-level activation peaks (see Section \ref{sec:BeatCondEmbed}) for $\{\mathbf{c}^\texttt{m}_t\}$ instead.
    \item \texttt{seq2seq+beat\,(high)}: using the beat condition embeddings from the high-level beat/downbeat positions as $\{\mathbf{c}^\texttt{m}_t\}$ instead.
    \item \texttt{seq2seq~w/o~beat}: to study the usefulness of the beat conditioning, we predicted $\{z_t^\texttt{d}\}$ from $\{z_t^\texttt{m}\}$, not using any beat-related conditions at all.
    \item \texttt{decoder+beat\,(low)}: to study the usefulness of the drumless codes $\{z_t^\texttt{m}\}$, we used only the low-level beat condition embeddings $\{\mathbf{c}^\texttt{m}_t\}$ as input to predict  $\{z_t^\texttt{d}\}$, via a Transformer decoder-only architecture (i.e., not seq2seq Transformer).
    \item \texttt{decoder~w/o~beat}: this is the baseline drummer that ``plays its own,'' generating $\{z_t^\texttt{d}\}$ autoregressively via a Transformer decoder without taking any information (neither $\{z_t^\texttt{m}\}$ nor $\{\mathbf{c}^\texttt{m}_t\}$) from the drumless clip $\mathbf{x}^\texttt{m}$ it is supposed to play along to.
\end{itemize}

\begin{table*}[]
\setlength{\tabcolsep}{4pt}
\centering
\begin{tabular}{|l|ccc|ccccc|}
\hline  
\multirow{2}{*}{Model} &
\multicolumn{3}{c|}{Objective metrics} & \multicolumn{5}{c|}{Subjective MOS ($\in [1,5]$)  } \\ 
& TrackEmd-MSE$\downarrow$ & Act-Entropy$\downarrow$  & B/DB-F1$\uparrow$  & ~~\textbf{R}$^\texttt{m/d}$~    & ~~\textbf{S}$^\texttt{m/d}$~    & ~~\textbf{Q}$^\texttt{d}$~    & ~~\textbf{R}$^\texttt{d}$~    & ~~\textbf{O}$^\texttt{d}$~    \\ \hline
\texttt{seq2seq+beat\,(low)}    & \textbf{.068}\small{$\pm.023$}   & \textbf{.928}\small{$\pm.189$} & \textbf{.340} & \textbf{3.27} & \textbf{3.44} & \textbf{3.39} & \textbf{3.37} & \textbf{3.20} \\
\texttt{seq2seq+beat\,(mid)}     & .077\small{$\pm.022$} & .963\small{$\pm.200$} & .212          & ---           & ---           & ---           & ---           & ---           \\
\texttt{seq2seq+beat\,(high)~}    & .080\small{$\pm.024$} & .938\small{$\pm.160$} & .132          & ---           & ---           & ---           & ---           & ---           \\
\texttt{seq2seq~w/o~beat}   & .081\small{$\pm.022$}   & .968\small{$\pm.233$}& .110          & 1.78          & 2.34          & 2.95          & 2.58          & 2.05          \\
\hline
\texttt{decoder+beat\,(low)} & \textbf{.068}\small{$\pm.023$}   & \textbf{.931}\small{$\pm.200$} & \textbf{.339} & \textbf{3.05} & \textbf{3.34} & \textbf{3.33} & \textbf{3.07} & \textbf{3.17} \\
\texttt{decoder~w/o~beat}   & .087\small{$\pm.025$}  & .987\small{$\pm.240$}  & .114          & 1.59          & 1.83          & 2.56          & 1.88          & 1.73             \\ \hline
Real data (not vocoded)    & ---   & --- & --- & 4.39 & 3.95 & 3.85 & 4.61 & 4.17 \\ 
\hline
\end{tabular}
\caption{Results of objective and subjective evaluation of  variants of the proposed JukeDrummer model. 
The metrics are (from left to right): beat/downbeat tracking embedding MSE, beat/downbeat activation entropy, beat/downbeat F1, \textbf{R}$^\texttt{m/d}$~(rhythmic consistency), \textbf{S}$^\texttt{m/d}$~(stylistic consistency), \textbf{Q}$^\texttt{d}$~(audio quality), \textbf{R}$^\texttt{d}$~(rhythmic stability), \textbf{O}$^\texttt{d}$~(overall). $\downarrow$\,/\,$\uparrow$: the lower/higher the better; best two results (among the six model variants) per column highlighted in bold.}
\label{tab:lm_config2}
\end{table*}

While the Mel-spectrogram for a clip in our case 
has 4,096 frames, the VQ-VAE encoder downsamples it to a sequence of $T=\text{1,024}$ latent vectors, namely each corresponding to $L=4$ frames.
For VQ-VAE, we set the codebook size of drumless sounds $K^\texttt{m}=\text{1,024}$ and that of drums $K^\texttt{d}=\text{32}$ (so $K^\texttt{m} \gg K^\texttt{d}$), and the dimension of the latent vectors $D=\text{64}$. 
The VQ-VAE encoders/decoders for both drumless and drums all have two layers.
For those models employing a seq2seq LM, the Transformer encoder has 9 layers and the Transformer decoder has 20 layers.\footnote{We used Adam as our optimizer with a learning rate of 0.0003 for both VQ-VAE and Transformer LM. We trained our LM with a batch size of 16, input feature dimension $d_\text{model}=\text{512}$, two heads for multi-head attention, and the chunk size for factorized attention being 16. For those employing a decoder-only architecture, the number of layers of the Transformer decoder reduces to 15 because 5 of the layers corresponding to the 
``encoder/decoder attention'' are removed.}

At inference time, we used the drum VQ decoder to convert the drum codes $\{z_t^\texttt{d}\}$ to a Mel-spectrogram, which is then turned into the waveform of the drum clip $\mathbf{x}^\texttt{d}$ by a HiFi-GAN V1 vocoder \cite{hifigan}. We trained the vocoder from scratch with audio of drum sounds from our  dataset for 2.5 days,
and then, inspired by \cite{assemvc22icassp,naturalspeech}, fine-tuned it on the re-constructed Mel-spectrograms of the Drum VQ decoder.
\section{Objective Evaluation}
\label{sec:exp}


As audio-domain drum accompaniment generation is new, we propose customized metrics. 
Specifically, we use the ``drum tracker''
trained exclusively for drum stems by Chiu \emph{et al.} \cite{spl21chiu}, which follows exactly the same  pipeline as the non-drum tracker (cf. Figure \ref{fig:beatinfo}), to extract rhythmic features at  three different levels for the ground-truth and generated drum sounds for the test split. We then compare the rhythmic features of the ground-truth and the generated in such three levels,
using the mean square error for the low-level tracker embeddings (\textbf{TrackEmb-MSE}), 
cross entropy for the mid-level activation functions (\textbf{Act-Entropy}), 
and the F-measure for beat/downbeat estimation (\textbf{B/DB-F1}).
The last one, computed by \texttt{mir\_eval} \cite{mireval},
uses the beat positions of the ground-truth drums as the reference and those of the generated drums as the estimated beats.  
These metrics evaluates only the \emph{rhythmic consistency} between the generated drums and the drumless audio (using its human-made drum track as a proxy), not other aspects such as stylistic consistency and audio quality, which will be evaluated subjectively  in Section \ref{sec:exp2}.

Result shown in the middle of Table \ref{tab:lm_config2} clearly shows that the models with low-level beat information achieve  lower MSE, cross entropy, and  higher F1, suggesting the usefulness of the low-level beat information for rhythmic consistency. The mid-level and high-level ones seem less effective (possibly because they are relatively monotonic), so we did not evaluate them further in Section \ref{sec:exp2}.
The objective scores also suggest that the Transformer encoder does not contribute much to rhythmic consistency.
\section{Subjective Evaluation}
\label{sec:exp2}

With an online study, we solicited 22 anonymous volunteers to rate the result for 3 out of 15 random drumless tracks (each 23.8 seconds)  from the test split.
Each time, a volunteer listened to a drumless tracks ($\mathbf{x}^\texttt{m}_*$) first, and then (in random orders) the mixture (i.e., $\mathbf{x}^\texttt{m}_*+\mathbf{x}^\texttt{d}_*$) containing drum samples generated by four different models, plus the real human-made one (to set a high anchor).
The volunteer then rated them in the following aspects on a 5-point Likert scale:
    \textbf{rhythmic consistency} between the drumless input and generated drums;
    \textbf{stylistic consistency} concerning the timbre and arrangement of the drumless input and generated drums;
    \textbf{audio quality} and \textbf{rhythmic stability} (whether the drummer follows a steady tempo) of the generated drum; and
    \textbf{overall} perceptual impression.

The mean opinion scores (MOS) in 
Table \ref{tab:lm_config2} 
show that $\texttt{seq2seq+beat\,(low)}$ consistently outperforms the others, validating the effectiveness of using both the drumless codes and beat conditions.
$\texttt{decoder+beat\,(low)}$ performs consistently the second best, outperforming the two models without beat information significantly in three aspects according to paired t-test  ($p$-value\,$<\text{0.05}$), validating again the importance of the beat-aware module.
Complementing Section \ref{sec:exp}, the MOS result suggests that the beat conditions seem more important than the drumless codes, though the best result is obtained with both.

\begin{figure}
    \centering
    \includegraphics[width=.81\linewidth]{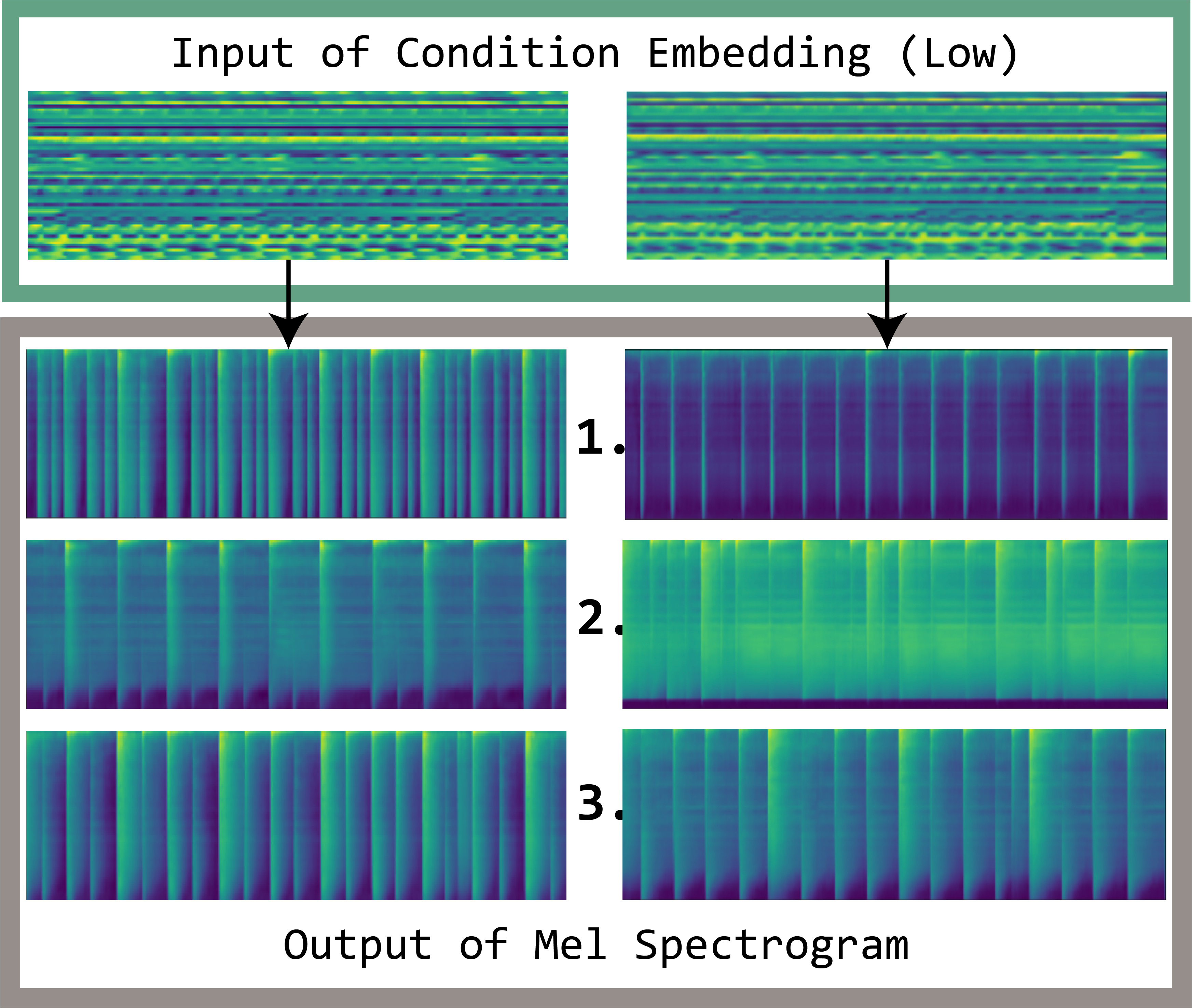}
    \caption{Two sets of three different generated samples by the same model given the same beat condition embedding.}
    \label{fig:diversity}
\end{figure}

Figure \ref{fig:diversity} further demonstrates that, given the same input, our model can generate multiple accompaniments with diversity in both beat and timbre.
Diversity is an interesting aspect that is hard to evaluate, but it is desirable as there is no single golden  drum accompaniment for a song. This may also explain why the F1 scores in Table \ref{tab:lm_config2} seem low.

Verbal feedbacks from the subjects confirm that our best model generates drum accompaniment that is rhythmically and stylistically consistent with the input, especially for band music or music with heavy use of bass.
However, the model still has limits. 
At times the model generates total silence, though it can be avoided by sampling the LM again.
The model may struggle to change its tempo going through different sections of a song.
Moreover, the generation might be out-of-sync with the input in the beginning few seconds, until the model gets sufficient context.
Please visit the demo page for various examples.

\vspace{-1mm}
\section{Conclusion}
We have presented JukeDrummer, a novel
audio-to-audio extension of  OpenAI's JukeBox model capable of adding the drum part of a drumfree recording in the audio domain.
To our knowledge, this represents the first attempt to audio-domain generation conditioned on drumless mixed audio.
With objective and subjective evaluations, we validated the effectiveness of the customized VQ-VAE plus the seq2seq Transformer design, and the proposed beat-aware module.
Among the beat conditions, we found that the low-level embeddings work the best.
Future work can be done to further improve the language model (LM), and to extend our work to other audio-to-audio generation tasks.









\section{Acknowledgement}
We are grateful to the anonymous reviewers for their valuable comments. This research is funded by grant NSTC 109-2628-E-001-002-MY2 from the National Science and Technology Council of Taiwan.

\bibliography{ISMIRtemplate}

\end{document}